\documentclass[cernpreprint, atlasdraft=false, USenglish, texmf, orcidlogo]{atlasdoc}
\usepackage{atlaspackage}
\usepackage{atlasbiblatex}

\usepackage{atlasphysics}
\graphicspath{{logos/}{figures/}}

\usepackage{ANA-HION-2024-01-PAPER-defs}

\AtlasTitle{Search for the jet-induced diffusion wake in the quark-gluon plasma via measurements of jet--track correlations in photon--jet events in Pb+Pb collisions at $\sqrt{\mathrm{s}_{\mathrm{NN}}}$ = 5.02 TeV with the ATLAS detector}

\AtlasAbstract{
This paper presents a measurement of jet--track correlations in photon--jet events, using 1.72~nb$^{-1}$ of Pb+Pb data
at $\sqrt{s_{\text{NN}}}$ = 5.02 TeV recorded with the ATLAS detector at the LHC.
Events with energetic photon--jet pairs are selected, where the photon and jet are approximately back-to-back in azimuth.
The angular correlation between jets and charged-particle tracks with transverse momentum ($p_\text{T}$) in the range 0.5--2.0 GeV in the hemisphere opposite to the jet, $|\Delta\phi\text{(jet,track)}|>\pi/2$, is measured as a function of their relative pseudorapidity difference, $|\Delta\eta\text{(jet,track)}|$.
In central Pb+Pb collisions, these correlations are predicted to be sensitive to the diffusion wake in the quark-gluon plasma resulting from the lost energy of high-$p_\text{T}$ partons traversing the plasma, with a characteristic modification as a function of $|\Delta\eta\text{(jet,track)}|$. The correlations are examined with different selections on the jet-to-photon $p_\text{T}$ ratio to select events with different degrees of energy loss.
No diffusion wake signal is observed within the current sensitivity and upper limits at 95\% confidence level on the diffusion wake amplitude are reported.
}

\AtlasRefCode{HION-2024-01}

\PreprintIdNumber{CERN-EP-2024-202}

\AtlasJournalRef{Phys. Rev. C 111 (2025) 044909}
\AtlasDOI{10.1103/PhysRevC.111.044909}

\hypersetup{pdftitle={ATLAS document},pdfauthor={The ATLAS Collaboration}}

\begin{document}

\maketitle
\newcommand{\AtlasCoordFootnote}{%
ATLAS uses a right-handed coordinate system with its origin at the nominal interaction point (IP) in the centre of the detector and the \(z\)-axis along the beam pipe. The \(x\)-axis points from the IP to the centre of the LHC ring, and the \(y\)-axis points upwards. Polar coordinates \((r,\phi)\) are used in the transverse plane, \(\phi\) being the azimuthal angle around the \(z\)-axis. The pseudorapidity is defined in terms of the polar angle \(\theta\) as \(\eta = -\ln \tan(\theta/2)\) and is equal to the rapidity $ y = \frac{1}{2} \ln \left( \frac{E + p_z c}{E - p_z c} \right) $ in the relativistic limit. Angular distance is measured in units of \(\Delta R \equiv \sqrt{(\Delta y)^{2} + (\Delta\phi)^{2}}\).}

\tableofcontents

\section{Introduction}
Collisions of high-energy nuclei at the Relativistic Heavy Ion Collider (RHIC) and the Large Hadron Collider (LHC) produce small droplets of quark-gluon plasma (QGP)~\cite{Elfner:2022iae}.   These QGP droplets quickly expand and are well described as a near-perfect (i.e., nearly inviscid) fluid~\cite{Romatschke:2007mq}.
One of the primary signatures of QGP formation is the substantial energy lost by large transverse momentum (\pt) quarks and gluons passing through the QGP.    This energy loss, often termed ``jet quenching'', indicates the presence of a medium with large color opacity~\cite{Cunqueiro:2021wls,Connors:2017ptx}.    Numerous measurements at the RHIC and at the LHC, when combined with theoretical predictions, enable the extraction of the total amount of energy lost by these partons while traversing the QGP.
This has been most recently achieved via the measurement of jet suppression in events tagged by prompt isolated photons, i.e., $\gamma$--jet observables~\cite{HION-2022-14}.

When a high-\pt parton loses energy, it is important to understand how that energy is distributed in terms of radiated gluons, i.e., what is the overall modification of the parton shower.  In addition, energy may also be transferred to the QGP fluid, which is well-described by hydrodynamic models with very small dissipation.    There are theoretical calculations predicting a ``medium response,'' including a Mach cone, a wake front (an enhanced amplitude of the medium in the direction of the parton), and an associated diffusion wake (a depletion in the amplitude of the medium in the opposite direction)~\cite{Casalderrey-Solana:2004fdk,Ruppert:2005uz,Betz:2008ka}. Measurements of this medium  response would provide important constraints on the speed of sound and viscosity of the QGP.   Many papers have detailed calculations of this medium response with different modeling of the lost energy and the QGP fluid itself~\cite{Chen:2017zte,Pablos:2022piv,Yang:2021qtl,Casalderrey-Solana:2020rsj,Brewer:2021hmh,Chen:2020tbl,Cao:2020wlm}.

As detailed in Ref.~\cite{Cao:2020wlm}, there are significant challenges to experimentally confirm these different medium response signatures.        The medium response in the direction of the parton competes with the modified parton shower and thus has not resulted in an unambiguous signature.    Observations of enhanced low-\pt particle production and particles at larger angles relative to the jet have been reported~\cite{Cao:2022odi,HION-2018-03,CMS:2018zze,STAR:2016jdz, PHENIX:2020alr}. However, their definitive attribution remains ambiguous.

In di-jet events, the diffusion wake (depletion) induced by one jet is contaminated from the wake (enhancement) of the other jet in the opposite direction, which may lead to a cancellation of observable effects. However, in $Z$/$\gamma$--jets events, the diffusion wake can be measured cleanly as $Z$/$\gamma$
do not interact strongly in the plasma and thus produce no medium response of their own.
Initial experimental results for $Z$--track correlations have been published~\cite{CMS:2021otx,HION-2019-05}, while Ref.~\cite{Cao:2020wlm} specifically proposes to search for the jet-induced diffusion wake in $Z$/$\gamma$--jets events.   In this paper, the higher statistics $\gamma$--jet channel  is pursued.

In Ref.~\cite{Yang:2022nei}, a new observable is suggested to aid in the separation of the medium response, in this case the diffusion wake, from other correlated particle production referred to as the multiparton interaction (MPI) contribution. Utilizing the Coupled Linear Boltzmann Transport and hydrodynamics (CoLBT-hydro) framework~\cite{Chen:2017zte}, a fully three-dimensional medium response can be mapped out.
This framework models $\gamma$--jet events and examines the correlation between the jet axis ($\eta_{\mathrm{jet}},\phi_{\mathrm{jet}}$)\footnote{\AtlasCoordFootnote} and low-\pt charged hadrons ($\eta_{\mathrm{h}},\phi_{\mathrm{h}}$). The absolute medium modification is then obtained by subtracting the correlation from the same hydrodynamic event without the $\gamma$--jet. The expected magnitude of the modification to the medium is of order 0.2\%~\cite{Yang:2022nei}.

The proposal in Ref.~\cite{Yang:2022nei} for separating the impacts of the diffusion wake and MPI is to examine this observable as a function of $\xjg = \ptj/\ptg$.
For events with lower $\xjg$, the quark or gluon opposing the photon loses more energy  on average in the medium and hence the diffusion wake is larger.     On the other hand, the MPI effect, being an initial-state effect, has no dependence on the energy loss effect, i.e., it is independent of \xjg.   Thus, testing the MPI independence on \xjg using proton--proton (\pppNoBlank) data is an important cross check.

This paper presents jet--track angular correlations utilizing $\gamma$--jet events in \pbpb collisions. The jet--track yield as a function of \detajh is compared to the one in events without the presence of a jet to extract the relative amount of diffusion wake compared to the bulk medium. This enables a direct test of these diffusion wake theory predictions.

\section{ATLAS detector}

The ATLAS detector~\cite{PERF-2007-01} at the LHC~\cite{Evans:2008zzb} covers nearly the entire solid angle around the collision point. %
It consists of an inner tracking detector surrounded by a thin superconducting solenoid, electromagnetic and hadronic calorimeters,
and a muon spectrometer incorporating three large superconducting air-core toroidal magnets.

The inner-detector system (ID) is immersed in a \qty{2}{\tesla} axial magnetic field
and provides charged-particle tracking in the range \(|\eta| < 2.5\).
The high-granularity silicon pixel detector covers the vertex region and typically provides four measurements per track,
the first hit generally being in the insertable B-layer (IBL) installed before Run~2~\cite{ATLAS-TDR-19,PIX-2018-001}.
It is followed by the SemiConductor Tracker (SCT), which usually provides eight measurements per track.
These silicon detectors are complemented by the transition radiation tracker (TRT),
which enables radially extended track reconstruction up to \(|\eta| = 2.0\).

The calorimeter system covers the pseudorapidity range \(|\eta| < 4.9\).
Within the region \(|\eta|< 3.2\), electromagnetic calorimetry is provided by barrel and
endcap high-granularity lead/liquid-argon (LAr) calorimeters,
with an additional thin LAr presampler covering \(|\eta| < 1.8\)
to correct for energy loss in material upstream of the calorimeters.
Hadronic calorimetry is provided by the steel/scintillator-tile calorimeter,
segmented into three barrel structures within \(|\eta| < 1.7\), and two copper/LAr hadronic endcap calorimeters.
The solid angle coverage is completed with forward copper/LAr and tungsten/LAr calorimeter modules
optimized for electromagnetic and hadronic energy measurements respectively.

Events are selected by the first-level trigger system implemented in custom hardware,
followed by selections made by algorithms implemented in software in the high-level trigger~\cite{TRIG-2016-01}.
The first-level trigger accepts events from the \qty{40}{\MHz} bunch crossings at a rate below \qty{100}{\kHz},
which the high-level trigger further reduces in order to record complete events to disk at about \qty{1}{\kHz}.

A software suite~\cite{SOFT-2022-02} is used in data simulation, in the reconstruction
and analysis of real and simulated data, in detector operations, and in the trigger and data acquisition
systems of the experiment.

\section{Event selections and simulations}

The datasets, photon and jet reconstruction, and simulation samples used in this measurement are identical to those used in a previous measurement of photon--tagged jet production~\cite{HION-2022-14}, and are briefly summarized here.

Events in data are selected for analysis using triggers requiring a reconstructed photon with transverse energy, $E_\mathrm{T}$, above $\text{35 GeV}$ (20 GeV) in \ppp (Pb+Pb) collisions~\cite{TRIG-2018-05}. These triggers sample the full luminosity of $255$~pb$^{-1}$ for the 2017 \ppp data and of $1.72$~nb$^{-1}$ for the 2018 \pbpb data, and are fully efficient for the photon selection used in this analysis.
In addition, minimum-bias (MB) triggered \pbpb events~\cite{HION-2020-10} are utilized for event-mixing as detailed below. Events are required to satisfy detector and data-quality requirements~\cite{DAPR-2018-01} and to have a reconstructed \ppp collision vertex from at least two tracks with $\pt > 500 ~\MeV$~\cite{ATL-PHYS-PUB-2015-026}. The vertex whose associated tracks give the highest sum of squared transverse momentum is designated the event primary vertex.
In \pbpb collisions, the forward calorimeters (FCal) covering $3.2 < \left|\eta\right| < 4.9$ is used to estimate the event centrality which is defined by the total transverse energy sum, $\Sigma{E}_\mathrm{T}^\mathrm{FCal}$.
Events in different intervals of $\Sigma{E}_\mathrm{T}^\mathrm{FCal}$ are associated with an underlying geometric configuration according to a Monte Carlo (MC) Glauber simulation~\cite{Miller:2007ri} using the same event selection criteria as in previous ATLAS analyses~\cite{HION-2016-07}.
This analysis uses a centrality interval corresponding to the 0--10\% of the $\Sigma{E}_\mathrm{T}^\mathrm{FCal}$ distribution in MB events.    This interval corresponds on average to the Pb+Pb collisions with the largest geometric overlap.

Simulated samples of $\gamma$--jet events, including direct and fragmentation photon contributions, were generated at leading order in QCD with \PYTHIA8~\cite{Sjostrand:2007gs} using the NNPDF2.3LO~\cite{Ball:2012cx} parton distribution function set and the A14~\cite{ATL-PHYS-PUB-2014-021} set of tuned parameters.
To include the effects of the underlying event (UE) in \pbpb collisions, the \PYTHIA8 $\gamma$--jet events are overlaid at the detector-hit level with \pbpb data recorded with minimum-bias triggers. These samples were simulated~\cite{SOFT-2010-01} using a \GEANT~\cite{Agostinelli:2002hh} description of the ATLAS detector and were digitized and reconstructed in a manner identical to that of the data.

Photons are reconstructed following the method used previously in Pb+Pb collisions~\cite{HION-2018-06,HION-2017-08}, which applies the procedure used in \ppp collisions~\cite{EGAM-2018-01} after an event-by-event estimation and subtraction of the UE contribution to the energy deposited in each calorimeter cell~\cite{HION-2011-02}. Photon candidates must pass shower shape requirements~\cite{ATLAS:2023dxj} designed to reject those arising from neutral meson decays and hadronic showers starting in the electromagnetic calorimeter.
Furthermore, photons are required to be isolated by requiring the sum of the transverse energy (after UE subtraction) in calorimeter cells within a cone of size $\Delta{R} = 0.3$ cone to be below optimized thresholds, achieving a 90\% efficiency for prompt photons in fine bins of centrality classes, as determined using the simulations described above. The photon isolation efficiency is evaluated with respect to generator-level final state photons which are isolated by requiring that the sum of the transverse energy of all the final-state particles, excluding the photon itself, within a cone of size $\Delta{R} = 0.4$ cone be less than 5~\GeV.

Jets are reconstructed following the procedure previously used in \pbpb collisions~\cite{HION-2011-02,HION-2017-10}. The anti-$k_t$ algorithm~\cite{Cacciari:2008gp,Fastjet} with distance parameter $R=0.4$ is applied to logical towers ($\Delta\eta\times\Delta\phi$ = $0.1\times\pi/32$), which are a combination of cells in all calorimeter layers. The contribution to the energy deposited in towers by the UE is estimated on an event-by-event basis, and the tower energies are iteratively updated to subtract the  UE contribution, which is then re-estimated. The resulting jets are corrected using simulation to account for the response of the calorimeter to jets~\cite{JETM-2018-05}, and then using {\em in situ} studies of jets recoiling against photons, $Z$ bosons, and jets in other regions of the calorimeter in \ppp collisions~\cite{HION-2018-06} for the absolute response in data. After performing this initial calibration, a process known as ``cross-calibration''~\cite{ATLAS-CONF-2015-016} is carried out. This step establishes a connection between the jet energy scale observed in high-luminosity \ppp collisions at $\sqrt{s}=13~\TeV$~\cite{ATLAS-CONF-2015-016} and the jets reconstructed using the different method described earlier in the 5.02 TeV \pbpb data.
The calibration described above is based on inclusive jets and an additional calibration correction is applied to account for the different flavor fraction estimated in the MC simulation between inclusive jets and jets produced in association with a photon.

Charged tracks are reconstructed following the procedure previously used in \pbpb collisions~\cite{HION-2011-03, ATL-PHYS-PUB-2015-051}. A selection criterion optimized for primary charged particles is used~\cite{HION-2017-04}.
Primary charged particles are defined as charged particles with a mean lifetime $\tau > 0.3 \times 10^{-10}$~s, either directly produced in the collision interactions or from subsequent decays of particles with a shorter lifetime~\cite{STDM-2010-06}.
All reconstructed tracks satisfying the selection criteria with $0.5 < \pt < 2.0$~\GeV and $|\eta| < 2.5$ are used in this analysis. This specific \pt range is selected because the medium response is expected to be most significant at lower transverse momenta. The charged-particle yield is corrected for reconstruction inefficiency, as well as tracks which are not associated with primary particles, on a per-track basis using simulation-derived correction factors.

\section{Analysis}
Events with photons passing the identification and isolation requirements described previously and with $ 90 < \etg < 180~\GeV$ and $\left|\eta^\gamma\right| < 2.37$ (excluding the region $1.37 < \left|\eta^\gamma\right| < 1.52$) are selected. Only the highest-\et (leading) photon among them is used in the measurement.
The kinematic selections of jets in this analysis are $\ptj > 40~\GeV$ and $\absetaj < 2.5$. These photon \etg and jet \ptj ranges encompass a broad range of \xjg, from 0.3 to 1.0.
The results are reported in three \xjg selections: \xjgthree, \xjgsix and \xjgeight. These \xjg ranges, by construction, impose upper \ptj limits corresponding to the photon \etg, which is restricted to below 180 GeV. The upper \etg boundary of $180~\GeV$ is imposed to facilitate comparison with the theoretical prediction.

The jet energy resolution (JER) and scale (JES) can lead to migration between \xjg ranges. However, this effect is found to be small and accounted for in the systematic uncertainty, so no unfolding is performed.
The azimuthal angle between the leading photon and associated jet, \dphigj, is required to be greater than $3\pi/4$. Only the leading jet in this \dphigj window is taken for the measurement.    These requirements
significantly reduce the rate of jets uncorrelated with the photon-producing hard scattering process as well as the contribution from multi-jet topologies.

For events with photon--jet pairs passing these selections, the distribution of the absolute pseudorapidity difference between the jet and each track, \detajh, is constructed.   All jet--track pairs must be in opposite azimuthal hemispheres, i.e., \dphijhabs > $\pi/2$. The \detajh distribution normalized by the number of photon--jet pairs is defined as
\begin{equation}
\raw = \frac{1}{N^\mathrm{\gamma-jet}} \frac{d^{2}N^\mathrm{jet-track}}{d\Delta\eta d\Delta\phi}.
\end{equation}

\begin{figure}[t!]
\begin{center}
\includegraphics[width=0.50\textwidth]{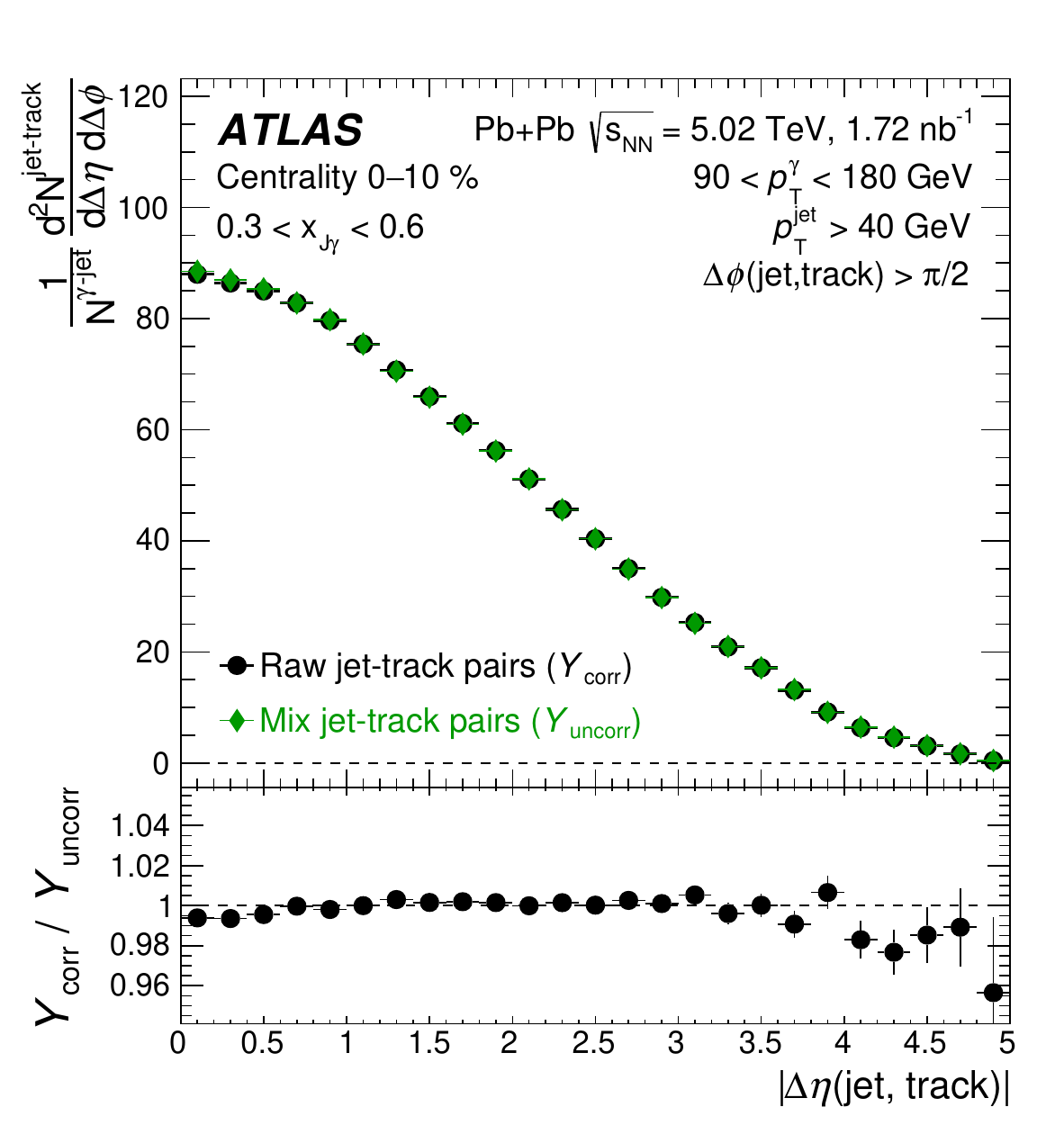}
\caption{
Top panel: the \detajh distributions for raw (\raw) and mixed (\mix) events for the \pbpb 0--10\% centrality interval for \xjgthree. Bottom panel: the ratio \rawtomix as a function of \detajh.  The vertical bars associated with symbols indicate the statistical uncertainties.
}
\label{fig:detajh_dist_Cent0to10}
\end{center}
\end{figure}

In \pbpb collisions, to gauge the medium modification of the QGP induced by the presence of jets, the tracks produced from the bulk medium constitute a background that is estimated using an event mixing technique and are used as a reference for the track--jet correlation in photon--jet events.
This ``uncorrelated'' track rate is estimated from the per-event track rate in MB \pbpb data.
A photon--jet pair in a given event is matched with tracks in a different  event, i.e., tracks from MB events that should have no $\textit{a~priori}$ relationship to a given photon--jet pair are used.
When mixing the two events, an MB Pb+Pb event is chosen to have similar properties as the signal event by matching %
\fcaletsum, the event plane angle~\cite{HION-2019-11}, and the $z$ position of the primary vertex.
In \pbpb collisions, the value of \fcaletsum in events with the photon--jet production (``signal'' event) includes a contribution from the photon--jet production and another one from the event without this photon--jet production. The \fcaletsum contribution from the photon--jet production is estimated in \ppp data, and has a mean value \fcaletsumpp = 17 GeV.  When mixing a signal event and an MB event, the \fcaletsum required is thus 17 GeV smaller than that of the signal event.
Figure~\ref{fig:detajh_dist_Cent0to10} shows the \detajh distributions from signal events (\raw) and from mixed events (labeled as \mix), and the ratio corresponding to \xjgthree. This ratio indicates the relative medium modification.

\begin{figure}[t!]
\begin{center}
\includegraphics[width=0.50\textwidth]{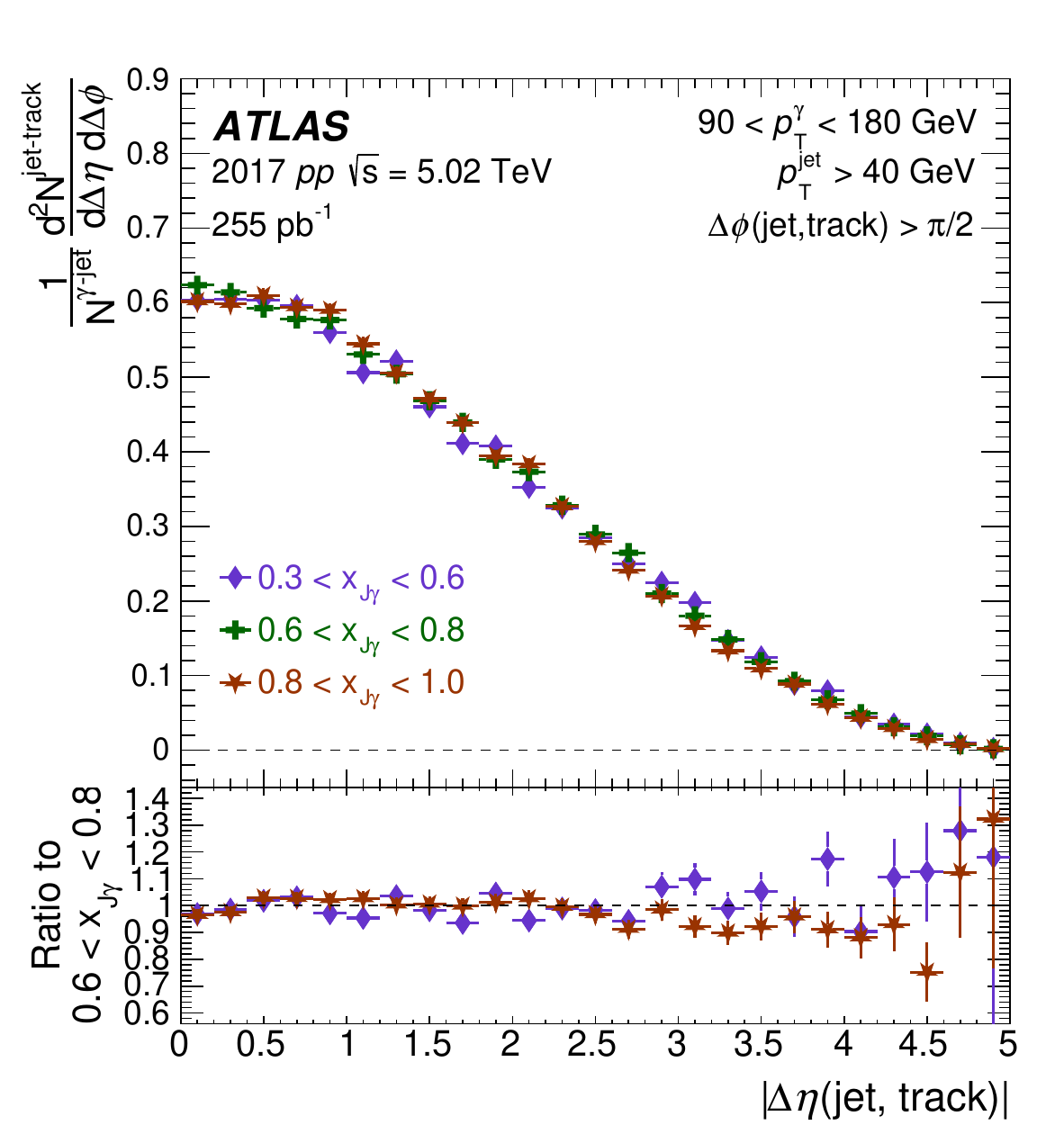}
\caption{Top panel: the raw \detajh distributions for different \xjg selections in \ppp collisions at 5.02~\TeV. Bottom panel: Ratio of yields in different \xjg selections to the one obtained for \xjgsix.     The vertical bars associated with each bin indicate the statistical uncertainties. }
\label{fig:detajh_dist_pp}
\end{center}
\end{figure}

As a check, the raw \detajh distributions in \ppp collisions at 5.02~\TeV are studied with the identical photon, jet, and track selections as in Pb+Pb collisions. In addition, the number of vertices is required to be exactly one to reject pileup events in \ppp collisions.
Figure~\ref{fig:detajh_dist_pp} shows the comparison
of the yield distributions of tracks per photon--jet pair as a function of \detajh in three selections of \xjg.
According to the theory expectations detailed in Ref.~\cite{Chen:2017zte}, the MPI should be independent of the specifics of the photon--jet kinematics. The presented ratio of the yields in different \xjg selections to the one obtained for \xjgsix is shown to be consistent with unity within statistical uncertainties in Figure~\ref{fig:detajh_dist_pp}, i.e., in agreement with the theoretical expectations.

\section{Systematic uncertainties}

The systematic uncertainties are evaluated by repeating the full analysis chain with a given systematic variation, which may result in, e.g., a different reconstructed-level distribution.
To avoid double-counting the statistical uncertainties, a \chisq\ test is performed for each source of systematic uncertainty.
Firstly, the signal samples are split into two halves for statistically independent comparisons between nominal and varied conditions: one half of the events for the nominal condition, the other half for the variation.
The \chisq\ of the difference between the variation and the nominal is calculated.
If the \chisq\ value is smaller than a threshold (\chisqcut), the differences are reasonably consistent within statistical fluctuations and thus no systematic uncertainty is assigned for this variation. The \chisqcut\ is set to correspond to the 68\% probability level, obtained by splitting the datasets 200 times under the same nominal condition, which reflects purely statistical fluctuations. Systematic sources which pass the \chisqcut\ are deemed systematically significant, whether due to a real systematic difference or as the result of a residual statistical fluctuation.
In this \chisq\ procedure, a small but real systematic difference may not be identified due to a statistical fluctuation in the nominal-variation event splitting.

\begin{figure}[t!]
\begin{center}
\includegraphics[width=0.98\textwidth]{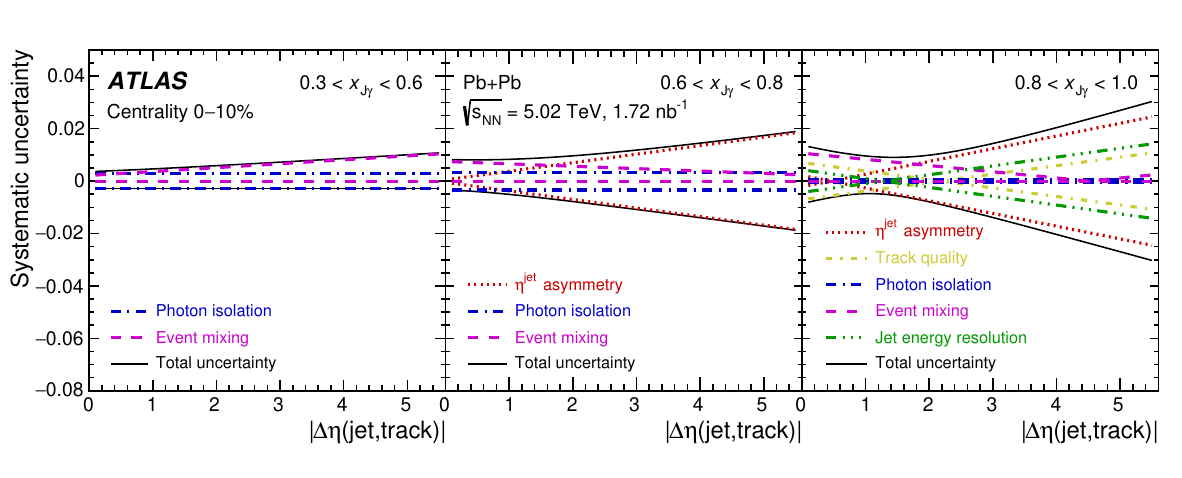}
\caption{Breakdown of the systematic uncertainties
as a function of \detajh for the \pbpb 0--10\% centrality interval. Different panels represent different \xjg ranges (\xjgthree, \xjgsix, and \xjgeight).}
\label{fig:syst}
\end{center}
\end{figure}

The sources of systematic uncertainty in this measurement are those associated with the track, jet, photon, and event mixing components. For track-related uncertainties, track selection criteria are varied using the same procedure as in Ref.~\cite{HION-2019-05}. Additionally, to account for the asymmetric detector performance in the ID, the analysis is repeated for \etatrk < 0 and \etatrk > 0, separately. Similarly, the \etaj asymmetry is considered as another source of systematic uncertainty arising from the imperfections in the calorimeter performance. In addition, the JER can shift jets between different \xjg selections. Therefore, the reconstructed jet \ptj is smeared using the JER for the variation.
The JES is also considered as a systematic uncertainty, but its effects are negligible.
Regarding photon-related uncertainties, a tighter photon isolation energy requirement is applied, setting the isolation threshold to achieve an 80\% isolation efficiency. For the nominal selection, the purity of isolated photons is high and there is thus no explicit correction made for background photons. The tighter isolation criterion is used to account for the impact of potential remaining background photons.
Also, to examine the impact of the photon isolation energy cone size (\dr = 0.3) on the results, the analysis is repeated with $\Delta\eta(\mathrm{jet, }\gamma) > 0.5$.
This variation thus excludes tracks that might directly influence the isolation energy calculation.
Finally, systematic uncertainties related to the event-mixing procedure are considered. The \fcaletsumpp estimation ($\text{17 GeV}$) is varied up and down by a conservative value of $\pm$50\%. For sources which have distinct ``up'' and ``down'' variations, i.e., event mixing, uncertainties are asymmetric.
For sources which only have a one-sided variation, uncertainties are symmetrized.

Figure~\ref{fig:syst} shows the breakdown of absolute systematic uncertainties for \rawtomix. All uncertainties are (partially) correlated between \raw and \mix, leading to relatively small systematic uncertainties in the ratio \rawtomix. Systematic uncertainty sources which fail the \chisq\ test are not depicted in the Figure and are not included as a contribution to the total uncertainty. As a result of the \chisq\ test procedure, different uncertainty sources may be included in the total uncertainty for the different \xjg ranges.
For \xjgthree, the dominant systematic uncertainty is the event-mixing uncertainty, and the total uncertainty ranges from smaller than 0.5\% at small \detajh to approximately 1\% at larger \detajh.
For \xjgeight, the total uncertainty increases from approximately 1\% at small \detajh (where the dominant contribution is the event-mixing component) to 2.5\% at large \detajh (where the dominant contribution is the \etaj asymmetry).

For the double ratio, \doubleratio, the different uncertainty contributions are evaluated according to the \chisq\ test specifically for this quantity by varying the numerator and denominator together.
Many uncertainty sources, such as track quality and jet energy resolution, are expected to have a similar impact in the different \xjg ranges and thus cancel in this double ratio. After evaluating these contributions, the photon isolation is the dominant remaining uncertainty in this measurement, with an approximately \detajh-independent magnitude of 0.5\%.

\section{Results}
\begin{figure}[t!]
\begin{center}
\includegraphics[width=0.98\textwidth]{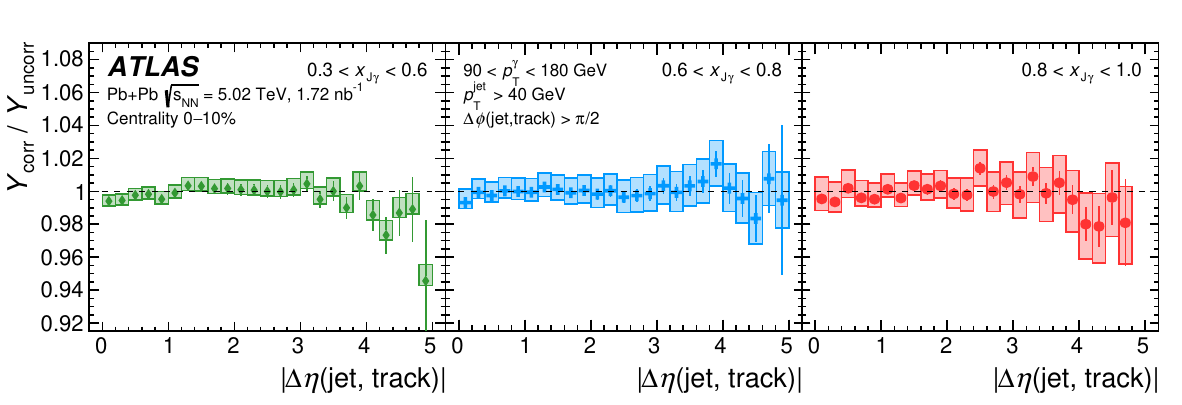}
\caption{
The \rawtomix distributions are shown as a function of \detajh for the \pbpb 0--10\% centrality interval. Different panels represent different \xjg ranges (\xjgthree, \xjgsix, and \xjgeight). The vertical bars %
indicate the statistical uncertainties.
The total systematic uncertainties are shown as boxes in each \detajh bin.}
\label{fig:res:rawtomix_Cent0to10}
\end{center}
\end{figure}

Figure~\ref{fig:res:rawtomix_Cent0to10} shows the ratio \rawtomix in the \pbpb 0--10\% centrality interval as a function of \detajh for jets and tracks in opposite azimuthal hemispheres ($\dphijhabs > \pi/2$), and in three categories of \xjg (\xjgthree, \xjgsix, \xjgeight). For all three \xjg selections, the results are consistent with unity within uncertainties.

\begin{figure}[t!]
\begin{center}
\includegraphics[width=0.49\textwidth]{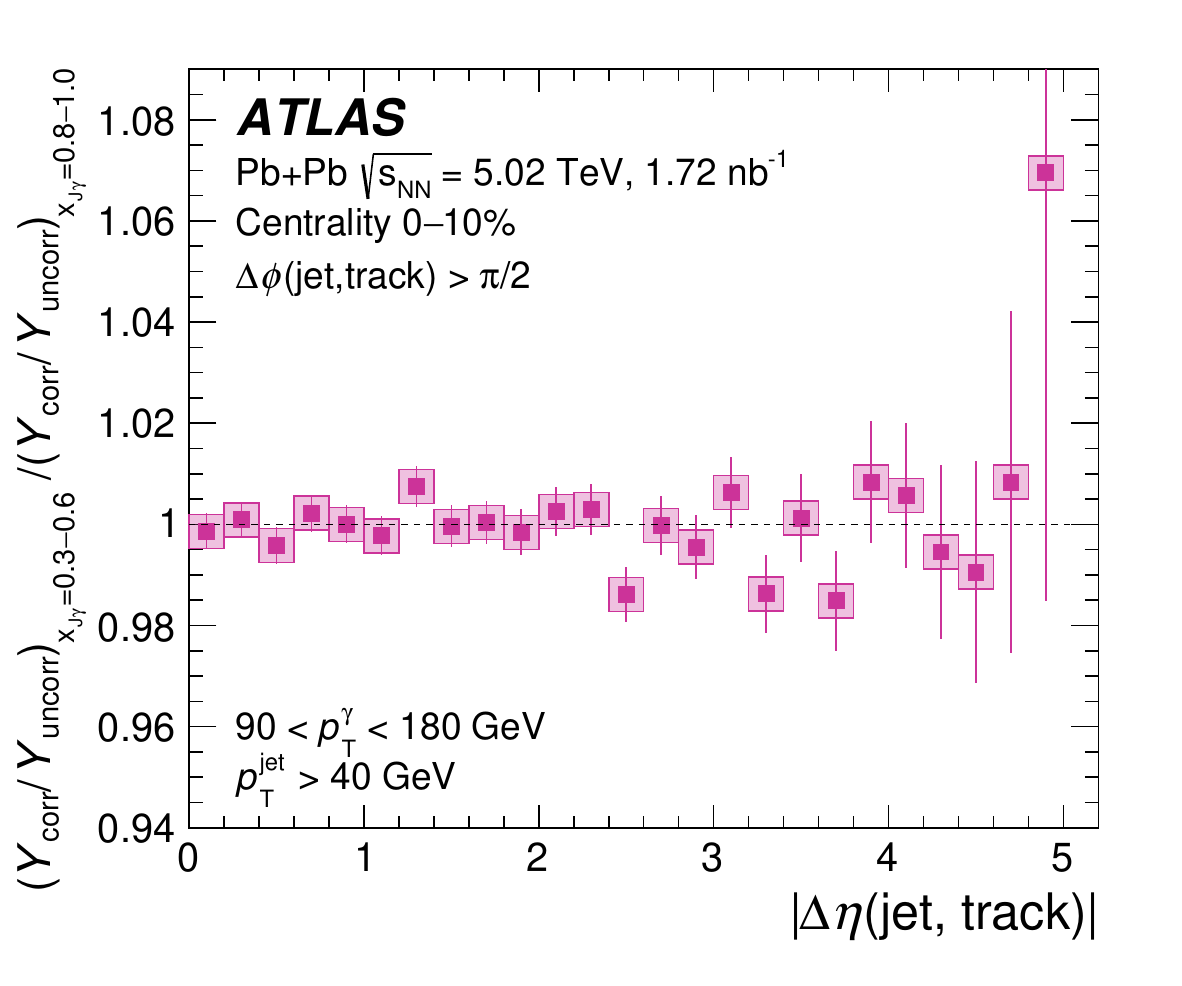}
\caption{
The double ratio \doubleratio as a function of \detajh for the \pbpb 0--10\% centrality interval.
The vertical bars indicate the statistical uncertainties.
The total systematic uncertainties are shown as boxes. }
\label{fig:res:doubleRatio_Cent0to10}
\end{center}
\end{figure}

Figure~\ref{fig:res:doubleRatio_Cent0to10} shows the double ratio, \doubleratio, which is particularly sensitive to whether a larger diffusion wake is present when the parton loses more energy in the QGP.   In addition, uncertainty sources that are correlated between \rawtomixthree and \rawtomixeight, e.g., event mixing uncertainties, partially cancel out in the ratio.
Again, the results are consistent with unity within uncertainties, meaning that no significant \xjg-dependence of the diffusion wake is found.

To quantify these observations further, the \rawtomix distributions are fitted with a function comprising a constant and a Gaussian term:
\begin{equation}
a_{0} + \adw e^{-\detajh^{2}/(2\sigmadw^2)},
\end{equation}
where \sigmadw and \adw correspond to the \detajh width and amplitude of the potential diffusion wake, respectively. The Gaussian shape of the diffusion wake is theoretically predicted. Such a diffusion wake would have a negative amplitude ($\adw < 0$).
For each value of \sigmadw, the most probable amplitude \adw is calculated via a MC sampling method including all statistical and systematic uncertainties and their correlations.
For input to theoretical models, it is convenient to calculate the best-fit \adw for different possible \sigmadw values. Thus, the fit is repeated with the \sigmadw parameter fixed, representing a different hypothesis each time, while \adw and $a_{0}$ are treated as free parameters.

\begin{figure}[t!]
\begin{center}
\includegraphics[width=0.98\textwidth]{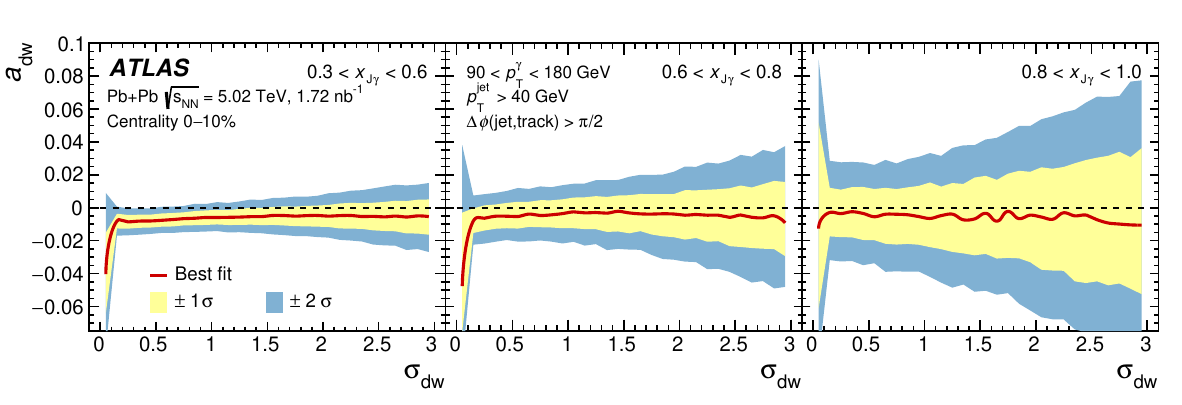}
\caption{
The diffusion wake amplitude \adw as a function of diffusion wake width \sigmadw from Gaussian fits for \rawtomix. Different panels represent different \xjg ranges (\xjgthree, \xjgsix, and \xjgeight). The red solid line is the most probable amplitude. The inner and outer shaded areas represent one ($\pm 1 \sigma$) and two ($\pm 2 \sigma$) standard deviation ranges, respectively.}
\label{fig:res:probFit_rawtomix_Cent0to10}
\end{center}
\end{figure}

Figure~\ref{fig:res:probFit_rawtomix_Cent0to10} shows the most probable values as well as the $\pm 1$ and $\pm 2$ standard deviation limits for the three \xjg selections.   For all diffusion wake widths \sigmadw and in all \xjg selections, the best-fit amplitudes are negative; however, all results are consistent with $\adw = 0$ (no signal), within approximately one (two) standard deviation for \xjgsix and \xjgeight (\xjgthree).
The systematic uncertainties between \detajh bins are highly correlated, and the statistical uncertainties dominate in the probability distributions of diffusion wake amplitudes.

\begin{figure}[t!]
\begin{center}
\includegraphics[width=0.49\textwidth]{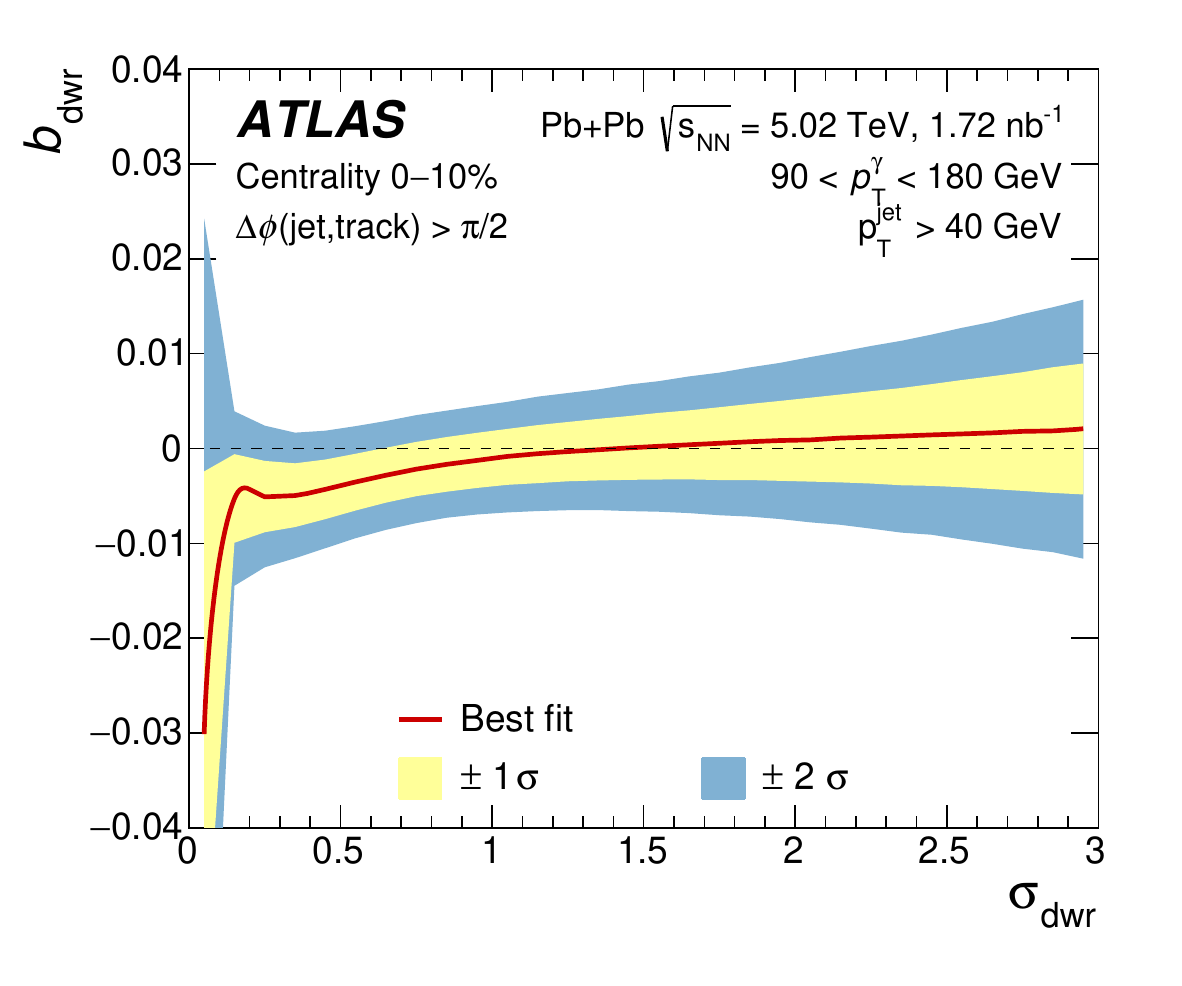}
\caption{
The amplitude (\bdwr) of a Gaussian fit for \doubleratio as a function of a given width \sigmadwr. The red solid line is the most probable amplitude. The inner and outer shaded areas represent one ($\pm 1 \sigma$) and two ($\pm 2 \sigma$) standard deviation ranges, respectively.}
\label{fig:res:probFit_doubleratio_Cent0to10}
\end{center}
\end{figure}

Similar to the \rawtomix fits, the \doubleratio distribution is fitted with the function defined as
\begin{equation}
b_{0} + \bdwr e^{-\detajh^{2}/(2\sigmadwr^2)} ,
\end{equation}
where \sigmadwr and \bdwr correspond to the \detajh width and amplitude of the double ratio, respectively. Figure~\ref{fig:res:probFit_doubleratio_Cent0to10} shows the best-fit relative diffusion wake amplitude (\bdwr) between \rawtomixthree and \rawtomixeight as a function of \sigmadwr (which is fixed in each fit). The most probable amplitude \bdwr is consistent with zero within one or two standard deviations, indicating that the best-fit amplitude \adw in \xjgeight is very similar to that in \xjgthree.   No significant diffusion wake signal that increases with larger parton energy loss is observed.

\begin{figure}[t!]
\begin{center}
\includegraphics[width=0.60\textwidth]{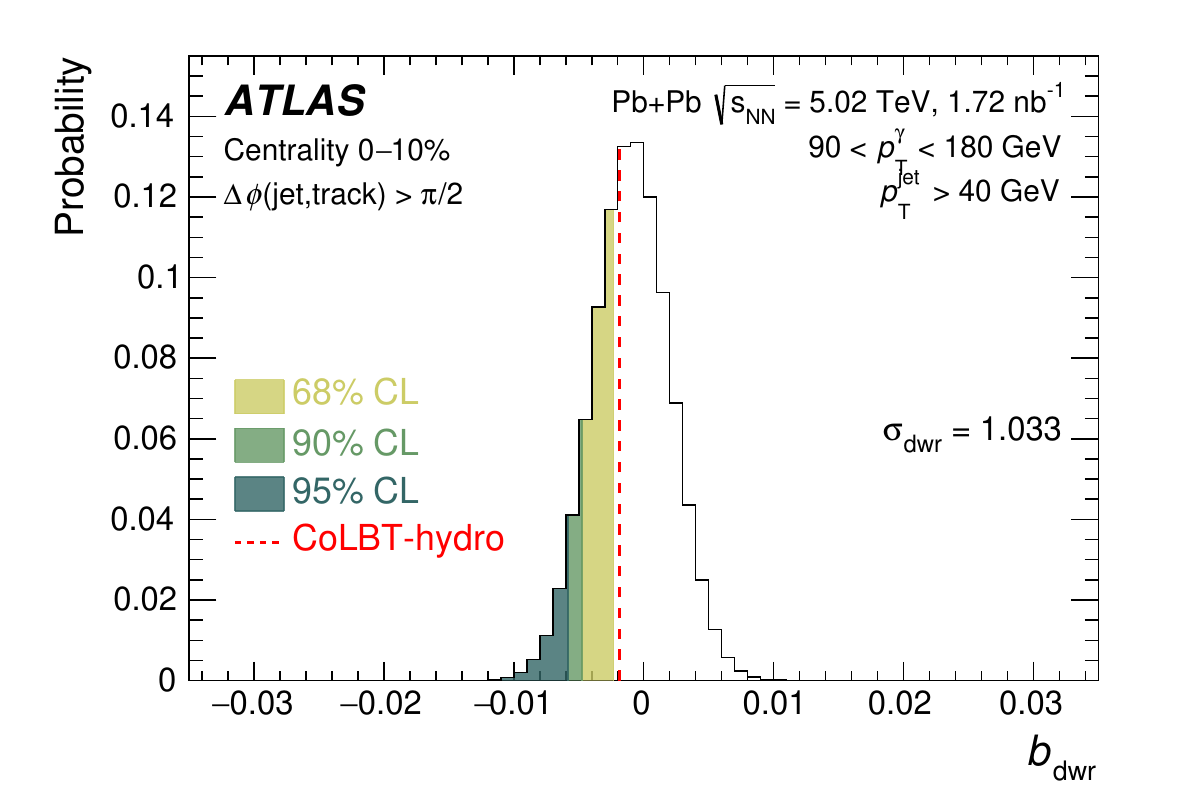}
\caption{
The probability distribution of the amplitude of the Gaussian fit component \bdwr for \doubleratio for \sigmadwr = 1.033 (the value predicted by the CoLBT-hydro framework). The yellow, green, and blue lines represent 68\%, 90\% and 95\% confidence levels, respectively.   The red dashed line is the theory expectation from the CoLBT-hydro framework. }
\label{fig:res:1D_probDist_theoryComp_width1p033}
\end{center}
\end{figure}

The theoretical framework CoLBT-hydro predicts a diffusion wake signal that increases for decreasing \xjg selections~\cite{Chen:2017zte}.    CoLBT-hydro calculations have been carried out to match the specific kinematic selections of this measurement, including the photon, jet, and track criteria.
The theory predicts diffusion wake parameters in the double ratio \doubleratio of $\bdwr = -0.00185$ and $\sigmadwr = 1.033$.
Figure~\ref{fig:res:1D_probDist_theoryComp_width1p033} shows the probability distribution for the double ratio when fixing \sigmadwr to 1.033.
The CoLBT-hydro theory expectation is overlaid. The small predicted \bdwr value is consistent with the experimental results within uncertainty. A diffusion wake double amplitude \bdwr value smaller than -0.0058 can be ruled out at 95\% confidence level.
The $p$-value for \bdwr being positive is 0.38.
As above, the constraining power of the measurement is limited by the statistical, rather than the systematic, precision of the dataset.

\FloatBarrier

\section{Conclusion}
The bulk quark-gluon plasma medium produced in heavy-ion collisions is expected to be modified by the energy lost from traversing jets. The expected localized depletion of the medium opposite to these jets is called the ``diffusion wake''. This paper presents measurements of angular correlations between jets and charged-particle tracks in photon--jet events, using 1.72~nb$^{-1}$ of Pb+Pb data at $\sqrt{s_{\mathrm{NN}}}$ = 5.02~\TeV recorded with the ATLAS detector at the LHC.
The measurement is performed for high-$p_\mathrm{T}$ photon--jet pairs, with three different ranges of the jet-to-photon $p_\mathrm{T}$ ratio, $x_\mathrm{J\gamma}$, intended to select events with different amounts of parton energy loss.
The yield of charged-particle tracks in the opposite hemisphere to the jet is measured as a function of the relative pseudorapidity separation, \detajh, and is divided by the yield of combinatoric jet--track pairs estimated in minimum-bias Pb+Pb events, to search for a localized depletion. The ratio of this ratio between different low and high $x_\mathrm{J\gamma}$ selections is also studied, and the probability of a diffusion signal with different parameters is estimated.
The data indicate no significant diffusion wake within the present uncertainties, which are dominated by the statistical uncertainties. The data are further used to set upper limits on the magnitude of the diffusion wake effect at different confidence levels.
The CoLBT-hydro theory prediction is consistent with the data within the 68\% confidence level upper limit. Assuming a double ratio width, \sigmadwr, given by the CoLBT-hydro model, values of the amplitude \bdwr smaller than -0.0058 are excluded at 95\% confidence level.
\section*{Acknowledgements}
%

%
%

%
%

We thank CERN for the very successful operation of the LHC and its injectors, as well as the support staff at
CERN and at our institutions worldwide without whom ATLAS could not be operated efficiently.

The crucial computing support from all WLCG partners is acknowledged gratefully, in particular from CERN, the ATLAS Tier-1 facilities at TRIUMF/SFU (Canada), NDGF (Denmark, Norway, Sweden), CC-IN2P3 (France), KIT/GridKA (Germany), INFN-CNAF (Italy), NL-T1 (Netherlands), PIC (Spain), RAL (UK) and BNL (USA), the Tier-2 facilities worldwide and large non-WLCG resource providers. Major contributors of computing resources are listed in Ref.~\cite{ATL-SOFT-PUB-2023-001}.

We gratefully acknowledge the support of ANPCyT, Argentina; YerPhI, Armenia; ARC, Australia; BMWFW and FWF, Austria; ANAS, Azerbaijan; CNPq and FAPESP, Brazil; NSERC, NRC and CFI, Canada; CERN; ANID, Chile; CAS, MOST and NSFC, China; Minciencias, Colombia; MEYS CR, Czech Republic; DNRF and DNSRC, Denmark; IN2P3-CNRS and CEA-DRF/IRFU, France; SRNSFG, Georgia; BMBF, HGF and MPG, Germany; GSRI, Greece; RGC and Hong Kong SAR, China; ISF and Benoziyo Center, Israel; INFN, Italy; MEXT and JSPS, Japan; CNRST, Morocco; NWO, Netherlands; RCN, Norway; MNiSW, Poland; FCT, Portugal; MNE/IFA, Romania; MSTDI, Serbia; MSSR, Slovakia; ARIS and MVZI, Slovenia; DSI/NRF, South Africa; MICIU/AEI, Spain; SRC and Wallenberg Foundation, Sweden; SERI, SNSF and Cantons of Bern and Geneva, Switzerland; NSTC, Taipei; TENMAK, T\"urkiye; STFC/UKRI, United Kingdom; DOE and NSF, United States of America.

Individual groups and members have received support from BCKDF, CANARIE, CRC and DRAC, Canada; CERN-CZ, FORTE and PRIMUS, Czech Republic; COST, ERC, ERDF, Horizon 2020, ICSC-NextGenerationEU and Marie Sk{\l}odowska-Curie Actions, European Union; Investissements d'Avenir Labex, Investissements d'Avenir Idex and ANR, France; DFG and AvH Foundation, Germany; Herakleitos, Thales and Aristeia programmes co-financed by EU-ESF and the Greek NSRF, Greece; BSF-NSF and MINERVA, Israel; NCN and NAWA, Poland; La Caixa Banking Foundation, CERCA Programme Generalitat de Catalunya and PROMETEO and GenT Programmes Generalitat Valenciana, Spain; G\"{o}ran Gustafssons Stiftelse, Sweden; The Royal Society and Leverhulme Trust, United Kingdom.

In addition, individual members wish to acknowledge support from Armenia: Yerevan Physics Institute (FAPERJ); CERN: European Organization for Nuclear Research (CERN PJAS); Chile: Agencia Nacional de Investigaci\'on y Desarrollo (FONDECYT 1230812, FONDECYT 1230987, FONDECYT 1240864); China: Chinese Ministry of Science and Technology (MOST-2023YFA1605700, MOST-2023YFA1609300), National Natural Science Foundation of China (NSFC - 12175119, NSFC 12275265, NSFC-12075060); Czech Republic: Czech Science Foundation (GACR - 24-11373S), Ministry of Education Youth and Sports (FORTE CZ.02.01.01/00/22\_008/0004632), PRIMUS Research Programme (PRIMUS/21/SCI/017); EU: H2020 European Research Council (ERC - 101002463); European Union: European Research Council (ERC - 948254, ERC 101089007), Horizon 2020 Framework Programme (MUCCA - CHIST-ERA-19-XAI-00), European Union, Future Artificial Intelligence Research (FAIR-NextGenerationEU PE00000013), Italian Center for High Performance Computing, Big Data and Quantum Computing (ICSC, NextGenerationEU); France: Agence Nationale de la Recherche (ANR-20-CE31-0013, ANR-21-CE31-0013, ANR-21-CE31-0022, ANR-22-EDIR-0002), Investissements d'Avenir Labex (ANR-11-LABX-0012); Germany: Baden-Württemberg Stiftung (BW Stiftung-Postdoc Eliteprogramme), Deutsche Forschungsgemeinschaft (DFG - 469666862, DFG - CR 312/5-2); Italy: Istituto Nazionale di Fisica Nucleare (ICSC, NextGenerationEU), Ministero dell'Università e della Ricerca (PRIN - 20223N7F8K - PNRR M4.C2.1.1); Japan: Japan Society for the Promotion of Science (JSPS KAKENHI JP22H01227, JSPS KAKENHI JP22H04944, JSPS KAKENHI JP22KK0227, JSPS KAKENHI JP23KK0245); Netherlands: Netherlands Organisation for Scientific Research (NWO Veni 2020 - VI.Veni.202.179); Norway: Research Council of Norway (RCN-314472); Poland: Ministry of Science and Higher Education (IDUB AGH, POB8, D4 no 9722), Polish National Agency for Academic Exchange (PPN/PPO/2020/1/00002/U/00001), Polish National Science Centre (NCN 2021/42/E/ST2/00350, NCN OPUS nr 2022/47/B/ST2/03059, NCN UMO-2019/34/E/ST2/00393, NCN \& H2020 MSCA 945339, UMO-2020/37/B/ST2/01043, UMO-2021/40/C/ST2/00187, UMO-2022/47/O/ST2/00148, UMO-2023/49/B/ST2/04085, UMO-2023/51/B/ST2/00920); Slovenia: Slovenian Research Agency (ARIS grant J1-3010); Spain: Generalitat Valenciana (Artemisa, FEDER, IDIFEDER/2018/048), Ministry of Science and Innovation (MCIN \& NextGenEU PCI2022-135018-2, MICIN \& FEDER PID2021-125273NB, RYC2019-028510-I, RYC2020-030254-I, RYC2021-031273-I, RYC2022-038164-I), PROMETEO and GenT Programmes Generalitat Valenciana (CIDEGENT/2019/027); Sweden: Carl Trygger Foundation (Carl Trygger Foundation CTS 22:2312), Swedish Research Council (Swedish Research Council 2023-04654, VR 2018-00482, VR 2022-03845, VR 2022-04683, VR 2023-03403, VR grant 2021-03651), Knut and Alice Wallenberg Foundation (KAW 2018.0157, KAW 2018.0458, KAW 2019.0447, KAW 2022.0358); Switzerland: Swiss National Science Foundation (SNSF - PCEFP2\_194658); United Kingdom: Leverhulme Trust (Leverhulme Trust RPG-2020-004), Royal Society (NIF-R1-231091); United States of America: U.S. Department of Energy (ECA DE-AC02-76SF00515), Neubauer Family Foundation.

%
%


%
\clearpage
\printbibliography
\clearpage
 
\begin{flushleft}
\hypersetup{urlcolor=black}
{\Large The ATLAS Collaboration}

\bigskip

\AtlasOrcid[0000-0002-6665-4934]{G.~Aad}$^\textrm{\scriptsize 104}$,
\AtlasOrcid[0000-0001-7616-1554]{E.~Aakvaag}$^\textrm{\scriptsize 17}$,
\AtlasOrcid[0000-0002-5888-2734]{B.~Abbott}$^\textrm{\scriptsize 123}$,
\AtlasOrcid[0000-0002-0287-5869]{S.~Abdelhameed}$^\textrm{\scriptsize 119a}$,
\AtlasOrcid[0000-0002-1002-1652]{K.~Abeling}$^\textrm{\scriptsize 56}$,
\AtlasOrcid[0000-0001-5763-2760]{N.J.~Abicht}$^\textrm{\scriptsize 50}$,
\AtlasOrcid[0000-0002-8496-9294]{S.H.~Abidi}$^\textrm{\scriptsize 30}$,
\AtlasOrcid[0009-0003-6578-220X]{M.~Aboelela}$^\textrm{\scriptsize 45}$,
\AtlasOrcid[0000-0002-9987-2292]{A.~Aboulhorma}$^\textrm{\scriptsize 36e}$,
\AtlasOrcid[0000-0001-5329-6640]{H.~Abramowicz}$^\textrm{\scriptsize 155}$
% ...
\bigskip
\\

% ...
$^{17}$Department for Physics and Technology, University of Bergen, Bergen; Norway.\\
% ...
$^{30}$Physics Department, Brookhaven National Laboratory, Upton NY; United States of America.\\
% ...
$^{36}$$^{(a)}$Facult\'e des Sciences Ain Chock, Universit\'e Hassan II de Casablanca;$^{(b)}$Facult\'{e} des Sciences, Universit\'{e} Ibn-Tofail, K\'{e}nitra;$^{(c)}$Facult\'e des Sciences Semlalia, Universit\'e Cadi Ayyad, LPHEA-Marrakech;$^{(d)}$LPMR, Facult\'e des Sciences, Universit\'e Mohamed Premier, Oujda;$^{(e)}$Facult\'e des sciences, Universit\'e Mohammed V, Rabat;$^{(f)}$Institute of Applied Physics, Mohammed VI Polytechnic University, Ben Guerir; Morocco.\\
% ...
$^{45}$Physics Department, Southern Methodist University, Dallas TX; United States of America.\\
% ...
$^{50}$Fakult\"{a}t Physik , Technische Universit{\"a}t Dortmund, Dortmund; Germany.\\
% ...
$^{56}$II. Physikalisches Institut, Georg-August-Universit\"{a}t G\"ottingen, G\"ottingen; Germany.\\
% ...
$^{104}$CPPM, Aix-Marseille Universit\'e, CNRS/IN2P3, Marseille; France.\\
% ...
$^{119}$$^{(a)}$New York University Abu Dhabi, Abu Dhabi;$^{(b)}$United Arab Emirates University, Al Ain; United Arab Emirates.\\
% ...
$^{123}$Homer L. Dodge Department of Physics and Astronomy, University of Oklahoma, Norman OK; United States of America.\\
% ...
$^{155}$Raymond and Beverly Sackler School of Physics and Astronomy, Tel Aviv University, Tel Aviv; Israel.\\
% ...
$^{a}$ Also Affiliated with an institute covered by a cooperation agreement with CERN.\\
$^{b}$ Also at An-Najah National University, Nablus; Palestine.\\
$^{c}$ Also at Borough of Manhattan Community College, City University of New York, New York NY; United States of America.\\
$^{d}$ Also at Center for High Energy Physics, Peking University; China.\\
$^{e}$ Also at Center for Interdisciplinary Research and Innovation (CIRI-AUTH), Thessaloniki; Greece.\\
$^{f}$ Also at CERN, Geneva; Switzerland.\\
% ...

\end{flushleft}


%

%
%

%

%
%
%
%

\end{document}